\documentclass{article}

\usepackage[preprint]{iaseai26}

\usepackage[utf8]{inputenc} % allow utf-8 input
\usepackage[T1]{fontenc}    % use 8-bit T1 fonts
\usepackage{url}            % simple URL typesetting
\usepackage{booktabs}       % professional-quality tables
\usepackage{threeparttable} % More tablesd
\usepackage{amsfonts}       % blackboard math symbols
\usepackage{nicefrac}       % compact symbols for 1/2, etc.
\usepackage{microtype}      % microtypography
\usepackage{xcolor}         % colors
\usepackage{array}
\usepackage{listings}

\usepackage{tabularx}
\usepackage{amsmath}        % Maths
\usepackage{amssymb}
\usepackage{graphicx}       % graphics
\usepackage{subcaption}  % For subfigures
\usepackage{multirow}
\usepackage{tcolorbox}
\usepackage{enumitem}
\setlist{nosep} % removes itemize extra spacing

\newcommand{\secref}[1]{\S~\ref{#1}}
\lstdefinelanguage{json}{
    basicstyle=\ttfamily\small,
    numbers=left,
    numberstyle=\tiny,
    stepnumber=1,
    numbersep=5pt,
    showstringspaces=false,
    breaklines=true,
    frame=single,
    literate=
     *{0}{{{\color{blue}0}}}{1}
      {1}{{{\color{blue}1}}}{1}
      {2}{{{\color{blue}2}}}{1}
      {3}{{{\color{blue}3}}}{1}
      {4}{{{\color{blue}4}}}{1}
      {5}{{{\color{blue}5}}}{1}
      {6}{{{\color{blue}6}}}{1}
      {7}{{{\color{blue}7}}}{1}
      {8}{{{\color{blue}8}}}{1}
      {9}{{{\color{blue}9}}}{1}
      {:}{{{\color{red}{:}}}}{1}
      {,}{{{\color{red}{,}}}}{1}
      {\{}{{{\color{black}{\{}}}}{1}
      {\}}{{{\color{black}{\}}}}}{1},
    string=[s]{"}{"},
    stringstyle=\color{teal},
    comment=[l]{//},
    commentstyle=\color{gray},
}

\title{Agent Benchmarks Fail Public Sector Requirements}

\author{
\textbf{Jonathan Rystrøm}$^{1,*}$ \quad
\textbf{Chris Schmitz}$^{2}$ \quad
\textbf{Karolina Korgul}$^{1}$ \quad
\textbf{Jan Batzner}$^{3,4}$ \And
\textbf{Chris Russell}$^{1}$ \\
\\
$^{1}$Oxford Internet Institute, University of Oxford, Oxford, United Kingdom \\
$^{2}$Centre for Digital Governance, Hertie School, Berlin, Germany \\
$^{3}$Weizenbaum Institute, Berlin, Germany \\
$^{4}$Technical University of Munich, Munich, Germany \\[0.25em]
\texttt{jonathan.rystrom@oii.ox.ac.uk}
}

\begin{document}    

\maketitle

\begin{abstract}
Deploying Large Language Model-based agents (LLM agents) in the public sector requires assuring that they meet the stringent legal, procedural, and structural requirements of public-sector institutions. Practitioners and researchers often turn to benchmarks for such assessments. However, it remains unclear what criteria benchmarks must meet to ensure they adequately reflect public-sector requirements, or how many existing benchmarks do so. In this paper, we first define such criteria based on a first-principles survey of public administration literature: benchmarks must be \emph{process-based}, \emph{realistic}, \emph{public-sector-specific} and report \emph{metrics} that reflect the unique requirements of the public sector. We analyse more than 1,300 benchmark papers for these criteria using an expert-validated LLM-assisted pipeline. Our results show that no single benchmark meets all of the criteria. Our findings provide a call to action for both researchers to develop public sector-relevant benchmarks and for public-sector officials to apply these criteria when evaluating their own agentic use cases.
\end{abstract}

\section{Introduction} \label{sec:intro}
In liberal democracies, public-sector organisations (PSOs) are expected to be neutral, efficient, unbiased entities faithfully enacting democratically negotiated laws \citep{bertelliAnchoringInstitutionalValues2025}. To ensure this, most jurisdictions place stringent requirements upon them: process legitimacy, political neutrality, transparency, and impersonality, to name a few \citep[see, e.g.,][]{grimmelikhuijsenLegitimacyAlgorithmicDecisionmaking2022,vanderwalGalaxiesUniverseCrossdisciplinary2015}. Breaches of these requirements can significantly undermine citizen trust in government \citep{lauxTrustworthyArtificialIntelligence2024}; where breaches become systematic or intentional, they may cause shifts in the ``narrow corridor'' balance between state and societal power \citep{acemogluNarrowCorridorStates2019}, threatening democratic stability.

The introduction of \emph{Large Language Model-based Agents (LLM Agents)} in public-sector institutions poses an unprecedented threat to these requirements. The promise (and peril) of LLM agents is their capability to accomplish more complex tasks with less oversight \citep{kasirzadehCharacterizingAIAgents2025}. For instance, LLM Agents could improve efficiency by assisting with compliance \citep{zhuComplianceBrainAssistant2025} or compiling evidence for complex policy documents \citep{musumeciLLMBasedMultiagent2024}. In both scenarios, LLM agents inherently make discretionary micro-decisions on which information to include and how to present it, which threatens the legitimacy and autonomy of the system \citep{schmitzMoralAgencyFramework2025}. In the long term, this discretion risks shifting power from human decision makers to unsteerable, opaque systems \citep{chanHarmsIncreasinglyAgentic2023}. Adapting these systems responsibly in public-sector contexts, therefore, requires robust guarantees that the resulting sociotechnical construct continues to comply with these uniquely strict requirements \citep{loweFullstackAlignmentCoaligning2025}. Such guarantees are enabled by reliable measurement of LLM agents' capabilities and risks \citep{schmitzOversightStructuresAgentic2025}.

The primary technical means for evaluation of model capabilities is \emph{benchmarking}.\footnote{Benchmarking is a component of larger oversight procedures; see \citet{mokanderAuditingLargeLanguage2024,schmitzOversightStructuresAgentic2025}.} Benchmarks are sets of tasks used to evaluate AI systems. While much machine learning research is structured around developing systems that perform well on the most popular benchmarks \citep{weidingerEvaluationScienceGenerative2025}, it is increasingly debated whether they provide much information of real-world use \citep{hardyMoreMarketingInformation2025}. Many benchmarks measure abstract capabilities on artificial tasks detached from the real-world contexts they aim to emulate \citep{rajiAIEverythingWhole2021,bowmanWhatWillIt2021}. This creates a \emph{benchmark-deployment} validity gap of an unknown size, which makes it difficult for practitioners across fields to make informed decisions. Public-sector deployments contend with a particularly severe form of this problem: it is unclear whether any given benchmark provides useful information, let alone certainty, on the conformity of an agentic system to public-sector requirements. 

\paragraph{Statement of Contributions:}
In this work, we build the foundation for bridging this gap. We do so by combining the literature on digitalisation and bureaucracy \citep[e.g.,][]{bullockArtificialIntelligenceBureaucratic2020} with theories of socio-technical evaluations of AI systems \citep{selbstFairnessAbstractionSociotechnical2019,wallachPositionEvaluatingGenerative2025}. This disconnect between current benchmarking practices and public-sector needs motivates our investigation into what constitutes appropriate evaluation for this domain.

To this end, we ask: 

\begin{enumerate}
    \item What criteria must a benchmark meet to guide safe and responsible adoption of LLM agents in the public sector?
    \item To what extent do current benchmarks for LLM agents meet these criteria?
\end{enumerate}

To answer the first question, we survey public administration literature to compile a list of operationalisable criteria against which to evaluate benchmarks (see \secref{sec:criteria}). We arrive at the following list: Benchmarks should be \emph{task-centric}; based on \emph{realistic} tasks and data; concerned with \emph{public sector-relevant} tasks; and report relevant \emph{metrics} including measures of \emph{cost} and \emph{fairness}. % and \emph{robustness}. 

Based on these theory-grounded criteria, we systematically analyse 1304 agentic LLM benchmark papers using a validated LLM-assisted pipeline (see \secref{sec:methods}). We show that no existing benchmark meets all criteria, with the biggest gaps existing for public-sector relevance and more comprehensive metrics (\secref{sec:results}). We suggest ways forward both for the technical design of public-sector agent benchmarks and the holistic sociotechnical evaluation of agent deployment in PSOs.

\section{Background}
\subsection{Agentic Large Language Models} \label{sec:agents-background}

\paragraph{Defining LLM Agents:} 
\citet{kasirzadehCharacterizingAIAgents2025} characterise AI agents as ``systems that have the ability to perform increasingly complex and impactful goal-directed action across multiple domains, with limited external control''. Their framework for understanding agentic LLMs consists of four key dimensions: autonomy, efficacy, goal complexity, and generality:

First, \textbf{autonomy} can be understood as the capacity to perform actions without external direction or control \citep{kasirzadehCharacterizingAIAgents2025}, with the key characteristic that they can ``take multiple consequential actions in rapid succession [...] before a human notices'' \citep{chanVisibilityAIAgents2024} without ``requiring step-by-step guidance'' \citep{kasirzadehCharacterizingAIAgents2025}. Second, \textbf{efficacy} refers to an agent's ability to directly interact with its environment \citep{kasirzadehCharacterizingAIAgents2025}. Particularly the ``access to external tools or services'' \citep{chanVisibilityAIAgents2024} and ``the ability to remove humans from the loop'' define agentic systems. Third, \textbf{goal complexity} represents agents' ability to form and pursue increasingly sophisticated goals, decomposing complex objectives into subgoals, and reliably adapt, plan, and execute over long time-horizons \citep{kasirzadehCharacterizingAIAgents2025,chanHarmsIncreasinglyAgentic2023,chanVisibilityAIAgents2024}. Finally, \textbf{generality} refers to an AI agent's ability to operate effectively across different roles, contexts, cognitive tasks, or ``economically valuable tasks'' \citep{chanVisibilityAIAgents2024}, ranging between ``highly specialised AI agents that are focused on specific tasks to general-purpose agents that can shift between different domains'' \citep{kasirzadehCharacterizingAIAgents2025}.

\paragraph{AI Agents and the Public Sector:} 
Public-sector tasks generally exhibit several features that make them amenable to automation: (a) clear documentation, (b) consistent structure, and (c) repetition \citep{schmitzOversightStructuresAgentic2025, bullockArtificialIntelligenceBureaucratic2020}. Previous scholarship has demonstrated the successes of LLM Agents on such narrowly defined and clearly documented tasks, like office management \citep{gurRealworldWebAgentPlanning2023}, manual healthcare delivery \citep{zhaoDualagentCollaborationFramework2025}, and software development \citep{jimenezSWEbenchCanLanguage2023}. However, what `success' means in the context of each paper is often poorly defined. We return to this question in \secref{sec:metrics}. 

Beyond narrow tasks, agents can act as interfaces and co-pilots. For instance, \citet{yunImprovingCitizengovernmentInteractions2024} argue that agents can help explain complex policies to citizens, and \citet[][from Meta]{zhuComplianceBrainAssistant2025} argue that LLM agents can assist with compliance. Again, whether these would work in practice requires strong benchmarks. We return to this in the next section. 

\subsection{Benchmarking}
\label{sec:benchmarking}

Benchmarks are commonly defined as standardised collections of \emph{datasets} (representing tasks or abilities) with \emph{scoring metrics}  \citep[representing performance on the task or ability;][]{rajiAIEverythingWhole2021,reuelBetterBenchAssessingAI2024}. Such benchmarks create a shared framework to compare different methods and measure progress. As such, benchmarks form the backbone of LLM evaluations.

% Benchmarks, which, following work by \citet{rajiAIEverythingWhole2021} and \citet{reuel2024betterbench} we define as standardised collections of datasets and a scoring metric, representing tasks or sets of abilities, which researchers chose as a shared framework to compare methods and performance, form the backbone of LLM evaluation. 

By operationalising an abstract capability (the phenomenon) into concrete problems and performance measures, benchmarks provide objective comparisons across architectures and training regimes \citep{bowmanWhatWillIt2021,beanMeasuringWhatMatters2025}. Crucially, a benchmark’s \textit{construct validity} -- the degree to which it truly measures the intended phenomenon -- determines whether high performance genuinely reflects a system's mastery of the underlying ability \citep{beanMeasuringWhatMatters2025}. For public-sector applications, where decisions may affect citizens’ rights and welfare, trusting benchmark scores without ensuring validity can evidently be dangerous \citep{rajiAIEverythingWhole2021}.

Drawing on psychometric principles, \citet{beanMeasuringWhatMatters2025} identify several essential features of a good benchmark:
\begin{itemize}
  \item[$\checkmark$] \textbf{Clear phenomenon definition:} Benchmarks must precisely define the targeted capability, whether ``multi‑step reasoning'', ``compliance checking'', or ``policy summarisation'', and, when relevant, decompose it into sub‑skills.
  \item[$\checkmark$] \textbf{Representative tasks:} Task sets should sample from realistic, domain‑relevant scenarios (e.g., authentic case documents or citizen queries), rather than rely solely on contrived or synthetic examples.
  \item[$\checkmark$] \textbf{Appropriate metrics:} Benchmarks should go beyond top-line measures like accuracy. We will discuss this in \secref{sec:metrics}.
  \item[$\checkmark$] \textbf{Rigorous methodology:} Robust statistical analyses, confidence intervals, significance tests, and contamination checks guard against overfitting to familiar data and ensure that improvements are meaningful rather than noise.
\end{itemize}

Unfortunately, many widely used benchmarks exhibit gaps. \textbf{Synthetic or toy tasks} are common, with nearly half of recent benchmarks augmenting small real-world cores with large volumes of LLM-generated examples, which risks inflated scores based on artefactual regularities \citep{beanMeasuringWhatMatters2025}. \textbf{Limited ecological validity} is another issue, as only about 30\% of benchmarks include evaluations in realistic simulators or field-like conditions, meaning that models may fail when confronted with the messiness of real public-sector workflows \citep{mcintoshInadequaciesLargeLanguage2025}. \textbf{Narrow metrics} further constrain interpretability, since over 80\,\% of benchmarks report only exact-match accuracy, omitting cost and other metrics critical for governmental use cases \citep{kapoorAIAgentsThat2025}. Finally, \textbf{sparse statistical rigour} undermines reliability, with fewer than 20\% of studies conducting statistical tests of differences, making it difficult to determine whether reported gains reflect genuine improvement or random fluctuation \citep{millerAddingErrorBars2024}.

% Unfortunately, many widely used benchmarks exhibit gaps.

% \textbf{Synthetic or toy tasks:} Nearly half of recent benchmarks augment small real-world cores with large volumes of LLM-generated examples, risking inflated scores on artefactual regularities \citep{beanMeasuringWhatMatters2025}.

% \textbf{Limited ecological validity:} Only about 30\% of benchmarks include evaluations in realistic simulators or field-like conditions, so models may fail when faced with the messiness of real public-sector workflows \citep{mcintoshInadequaciesLargeLanguage2025}.

% \textbf{Narrow metrics:} Over 80\,\% of benchmarks report only exact-match accuracy, omitting cost and other metrics that are critical for governmental use cases \citep{kapoorAIAgentsThat2025}.

% \textbf{Sparse statistical rigour:} Fewer than 20\% of studies conduct statistical testing of differences, making it hard to judge whether reported gains reflect genuine advancement or random fluctuation \citep{millerAddingErrorBars2024}.

For public institutions, these deficiencies translate into \emph{deployment risk}: an agent that ``succeeds'' on a simplistic benchmark may flounder when processing ambiguous citizen correspondence, exhibit brittleness to phrasing variations, or disproportionately disadvantage vulnerable groups. To mitigate these risks, future benchmarks for LLM‑based agents in public‑sector contexts should adhere to the construct‑validity framework, including defining phenomena precisely, sampling tasks representatively, reporting multi‑dimensional metrics, and applying rigorous statistical methods \citep{beanMeasuringWhatMatters2025,weidingerEvaluationScienceGenerative2025}. Only by ``measuring what matters'' can policymakers gain the clear‑eyed insights needed to deploy LLM agents responsibly in service of the public good. 

% The same is true of the applicability of these benchmarks in social-scientific research, including that on the impact of AI in PSOs. Well-established theories such as screen- and system-level bureaucracy \citep{bovensStreetlevelSystemlevelBureaucracies2002}, artificial discretion \citep{vredenburghAIBureaucraticDiscretion2025}, and diffusion of innovations \citep{rogersDiffusionInnovations2014} have acted as lenses for the study of the personal and organisational impacts of public-sector AI adoption \citep[see, e.g.,][]{madanAIAdoptionDiffusion2023,cetinapresuelAdoptionArtificialIntelligence2024}. Technical benchmarks have had little role in the generation of these insights. Work in these fields has instead largely relied on case studies of implementations \citep{daviesMoveFastBreak2025}, experiments \citep{seltenJustThoughtStreetlevel2023,konigHowGovernmentUses2025}, and surveys and interviews \citep{schmitzOversightStructuresAgentic2025}. The disconnect between benchmarks and theory is surprising, since benchmarks could help ground the theories in an empirical understanding of the technical frontier.

%Given the readily apparent value a well-founded understanding of the technical frontier would have for these theories, this is surprising.

What explains the disconnect between public-sector requirements and benchmarking practice? We hypothesise that it is owed to the design of these benchmarks, which is often driven by technical ``paths of least resistance'', conventions in computer science, and data availability, rather than their usefulness for real-world applications \citep{weidingerEvaluationScienceGenerative2025}. Benchmarks provide static measurements of the capacity of models to perform certain tasks that neither meaningfully advance nor test these hypotheses relevant to real-world use \citep{hardyMoreMarketingInformation2025}.

\section{Criteria for a Good Public-Sector Benchmark} \label{sec:criteria}

In this section, we concretise the notion of a ``good'' public-sector agent benchmark in the form of six criteria. As established above, the increasing benchmark-deployment validity gap is casting attention on the ecological validity, or real-world representativeness, of LLM and agent benchmarks. While the problems with existing benchmarks are well-mapped \citep{erikssonCanWeTrust2025}, there are few concrete characterisations of benchmarks that remedy these issues. As a result, the usefulness of benchmarks for tasks beyond progress measurement, such as for industry practitioners deciding on models, is often limited \citep{hardyMoreMarketingInformation2025}.

Accordingly, this section moves in the other direction. It derives a set of criteria for a ``good" public-sector agent benchmark from theoretical work across public administration, theories of automation, and technical benchmarking literature. As posed in RQ1, we aim to find criteria that make benchmarks theoretically and practically meaningful for the public sector. In describing these ideal-type criteria, we do not consider the effort required or even the technical feasibility of such a benchmark.

Public-admininstration theories, which are often descriptive or even normative,  should not be treated as blueprints to be reproduced or as claims to be empirically validated through benchmarking. Rather, their use in this exercise is that they produce intermediate abstractions which plausibly generalize across public-sector contexts. Our objective is to make use of these abstractions to strengthen the context-specific \textit{construct validity} of agent benchmarks. As such, we do not aim to set an upper bound on benchmark quality -- which is likely to remain case- and deployment-specific -- but to raise the floor of what qualifies as informative benchmarks in a public-sector context.

To select relevant social scientific theories, we adopted a purposeful sampling approach \citep{palinkasPurposefulSamplingQualitative2015}, selecting for theories that are (i) well-established and notable, (ii) salient in research on public-sector AI use, and (iii) robust and distinct. This resulted in a list of seven theoretical lenses, which we synthesised into six benchmark criteria, grouped into two categories: requirements for the structure of a benchmark, and requirements for the quantitative metrics the benchmark reports. The process is summarised in Fig. \ref{fig:process-overview} and detailed in the appendix, and the remainder of this section introduces and justifies these criteria.

\begin{figure}
    \centering
    \includegraphics[width=1\linewidth]{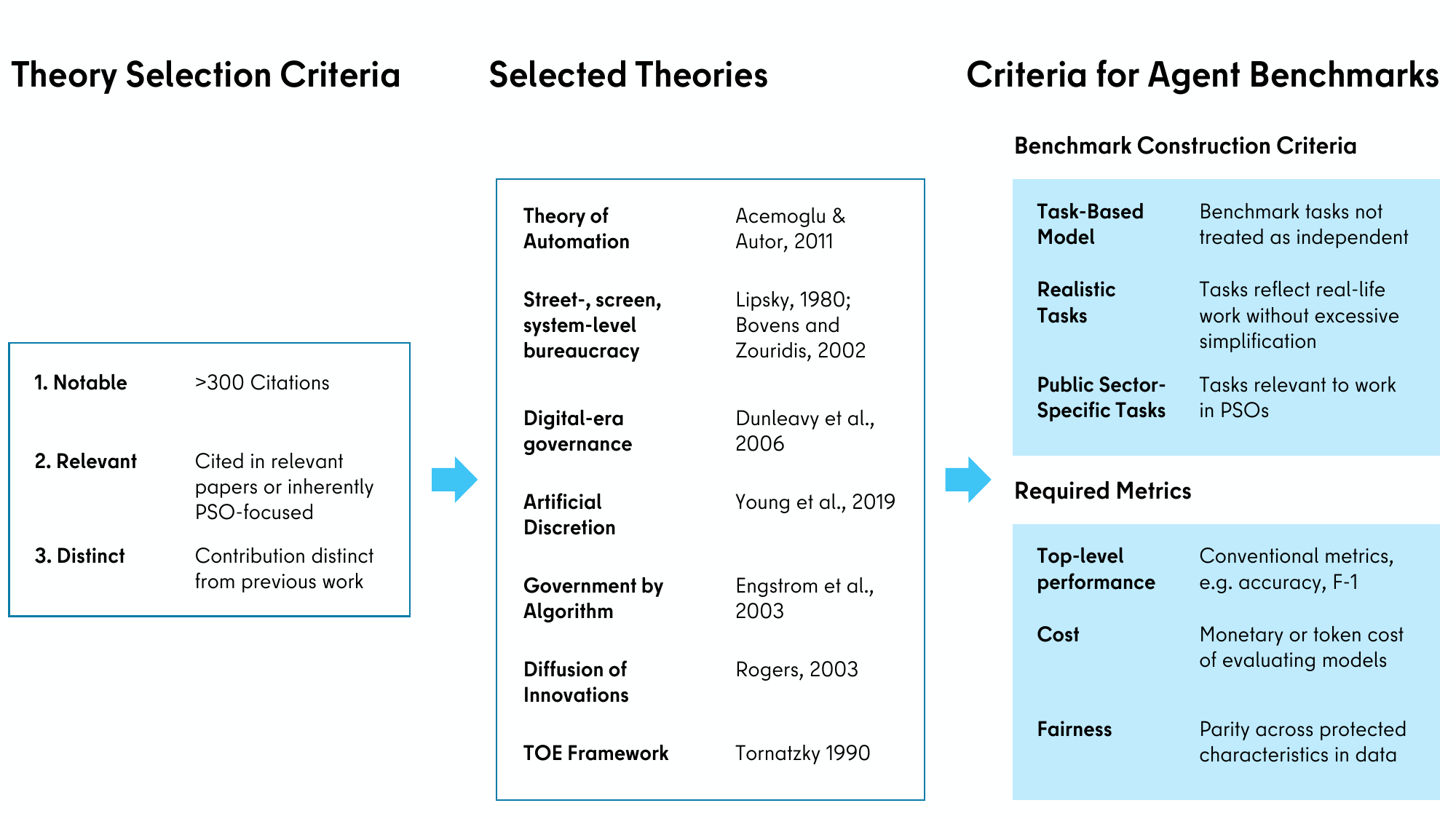}
    \caption{Overview of process for formulation of public sector-relevant agent benchmark criteria.}
    \label{fig:process-overview}
\end{figure}

\subsection{Benchmark Construction}

\paragraph{Task-based Model:}

The most obvious link between the studied theories is the conceptual deconstruction of organisations, systems, and processes into their constituent \emph{tasks}. Tasks are often defined as ``the smallest unit of activity with a meaningful outcome'' \citep[e.g., by the widely-used O*NET database;][]{dierdorffSummaryProceduresONET2011}. In automation theory, for example, \citet{acemogluSkillsTasksTechnologies2011} describe tasks as the ``fundamental unit of work''. \citet{bovensStreetlevelSystemlevelBureaucracies2002} describe the incremental task-level digitisation of public-sector processes as the vector by which the transition to screen- and system-level bureaucracy occurs, e.g., the entry of forms by public officials. The digital-era governance framework rests deeply on modularity, both for its ``do once'' design principle and the end-to-end digitalisation of processes \citep{dunleavyDesignPrinciplesEssentially2015}. \citet{rogersDiffusionInnovations2014} does not introduce a task-based lens to the diffusion of innovations, but does refer to this same modularity as an enabler of the iterative processes of innovation and organisation adaption that enable diffusion. Literature on artificial discretion and decision-making in PSOs frequently isolates discretionary and administrative tasks \citep{youngArtificialDiscretionTool2019, seltenJustThoughtStreetlevel2023}. Practical work applying these theories to AI, including in the public sector, also makes use of this decomposition \citep{straubAIBureaucraticProductivity2024, hashemMappingPotentialGenerative2025,gmyrekGenerativeAIJobs2023}. 

The value of these theories is in the re-aggregation of these tasks into superstructures. \citet{acemogluSkillsTasksTechnologies2011} aggregate tasks into \emph{jobs} to enable task-based measurement of automation exposure. Both the concepts of screen- and system-level bureaucracy \citep{bovensStreetlevelSystemlevelBureaucracies2002} and digital-era governance \citep{dunleavyDigitalEraGovernance2006} aggregate tasks into \emph{processes}. This enables (i) the introduction of dependence relationships between tasks, where the outputs of one task are required for subsequent ones, and (ii) the modular reuse and integration of tasks and their outputs across workflows.

A good benchmark should therefore:
\begin{itemize}
    \item Evaluate a model or system on individually meaningful tasks or units.
    \item Aggregate these tasks into structures, such as processes, which create dependence relationships between them.
\end{itemize}

An example of a good process-based paper is AutoPenBench \citep{gioacchiniAutoPenBenchVulnerabilityTesting2025}, which studies penetration testing through five well-defined, interconnected stages from discovery through to flag capturing.
\paragraph{Realistic Tasks:}

Each studied theory presupposes, frequently implicitly, that its composite tasks are actual pieces of work that are performed in its respective context. Variations of this requirement are readily apparent: diffusion of innovations \citep{rogersDiffusionInnovations2014} defines an innovation's relative advantage as its ability to fulfil real-world needs; automation theory requires tasks to be value-producing \citep{acemogluSkillsTasksTechnologies2011}; literature on AI adoption and government by algorithm categorizes AI use cases in terms of concrete previously human-performed tasks \citep{engstromGovernmentAlgorithmArtificial2020, daubAutomationGovernmentHarnessing2020, neumannExploringArtificialIntelligence2024}. 

As several sources show, this assumption is frequently broken by benchmarks which, for example, use machine-generated or isolated questions, mock or simplify environments, or ignore real-life contextual complexity \citep{hardyMoreMarketingInformation2025,selbstFairnessAbstractionSociotechnical2019}. 

Beyond the realism of individual tasks, studied theories presuppose that the \emph{collective} of all tasks is itself a realistic representation of the unit being studied, such as an occupation or a process. Two variations of this emerge. First, the distribution of tasks being studied must reflect the real-life distribution of tasks. For example, work on algorithmic discretion frequently highlights that discretionary decision-making itself requires only a small fraction of a public servant's time, compared to administrative and coordinative tasks \citep{bullockArtificialIntelligenceBureaucratic2020}. Second, the totality of tasks must indeed constitute the entire aggregate unit -- with, e.g. no coordinative, translational, or administrative steps skipped in the reconstruction. This is prominent in literature reviewing past failures of algorithmic governance, such as the Dutch benefits or Michigan MIDAS cases, which frequently point to an overemphasis on the decision-making steps of processes over the preparatory and subsequent ones \citep{peetersAdministrativeExclusionInfrastructurelevel2023, elyounesComputerSaysNo2021}.

A good benchmark should therefore:
\begin{itemize}
    \item Ensure the internal realism of each evaluated task, such as the context, tools, required output, and constraints. 
    \item Ensure the totality of tasks tested captures the entirety of the phenomenon being evaluated in a representative distribution.
\end{itemize}

A good example of a realistic benchmark is \citet{chenScienceAgentBenchRigorousAssessment2024}, which studies end-to-end scientific data analysis. It includes real, expert validated tasks.
\paragraph{Public Sector-Specific Tasks:}

% Of the studied theories, those specific to the public sector are reflective of the unique conditions of PSOs, which are overwhelmingly Weberian bureaucracies. In fulfillment of what \citet{gerthMaxWeberEssays1997} term \emph{bureaucratic rationality}, PSOs are subject to strict requirements of process justice, impartiality and impersonality, and documented procedure, to name a few. Street-level bureaucracy \citep{lipskyStreetlevelBureaucracy30th2010} is a formulation of how individual agents can fulfil these requirements under uncertain conditions. Digital-era governance \citep{dunleavyDigitalEraGovernance2006} phrases digitalisation as an enabler of transparency and auditability. Most centrally, government by algorithm \citep{engstromGovernmentAlgorithmArtificial2020} is a continuation of the hypothesis posed by \citet{simonAdministrativeBehaviorStudy1947} that the above conditions are best-met by digital systems. Independent of the normative ideal, the resulting clarity of process, documentation, and precedent constitutes technically ideal conditions for implementation of agentic systems \citep{schmitzOversightStructuresAgentic2025}. 

The studied theories reflect the unique conditions of PSOs, which are overwhelmingly Weberian bureaucracies \citep{gerthMaxWeberEssays1997}. Though PSOs can be structurally similar to organisations in the private sector and perform comparable tasks \citep{raineyComparingPublicPrivate2000}, they work under exceptionally strict requirements of process justice and documentation \citep{grimmelikhuijsenLegitimacyAlgorithmicDecisionmaking2022} upheld, for example, by strong traditions of administrative law encoding public values \citep{brysonPublicValueGovernance2014}. These constraints can produce a meaningfully different ``jagged frontier'' of AI performance \citep{dellacquaNavigatingJaggedTechnological2023}, even though the well-documented processes and strong regularization may in theory create \emph{more} favourable conditions for agentic automation and oversight \citep{schmitzOversightStructuresAgentic2025}. As such, the complex interactions between system-level rules and street-level bureaucracy create unique conditions which require explicit study \citep{lipskyStreetlevelBureaucracy30th2010, alkhatibStreetLevelAlgorithmsTheory2019}.

A good benchmark should therefore measure performance on tasks that either specifically occur in the public sector, or underlie the same requirements. A good example is AgentsCourt \citep{heAgentsCourtBuildingJudicial2024}, which measures judicial decision-making.

\subsection{Reporting relevant metrics} \label{sec:metrics}
Useful benchmarks require useful metrics. Metrics define \emph{what} agents are evaluated on \citep{beanMeasuringWhatMatters2025}. While all benchmarks report at least one top-level performance metric (e.g., `Success rate' or `Accuracy'), this is insufficient for making decisions about whether to implement an agent to solve a specific task \citep{meimandiMeasurementImbalanceAgentic2025}. All public-sector applications must meet statutory requirements set by the law \citep{wachterWhyFairnessCannot2021} as well as expectations from citizens and politicians \citep{grimmelikhuijsenLegitimacyAlgorithmicDecisionmaking2022}. As a minimum to meet these requirements, we argue that PSO-relevant benchmarks must also report sector-relevant metrics of the following: \textbf{Cost} and \textbf{Fairness}. We argue for these below:  %\textbf{Robustness}, and 

\paragraph{Cost:}
The first additional metric to define is how expensive an agent system is to run, i.e., the \emph{cost}. As shown by \citet{kapoorAIAgentsThat2025}, the cost of different agent systems can differ by orders of magnitude. The cost of an agent system has two components: 1) the cost per token, and 2) the number of tokens required per task. The cost per token principally depends on the size of the underlying LLM \citep{samsiWordsWattsBenchmarking2023}. Crucially, this represents a trade-off: larger models are often more capable and can thus solve a given task in fewer tokens \citep{wuInferenceScalingLaws2024}. This trade-off can be analysed by showing the \emph{Pareto curve} of cost versus performance.

Cost is an essential component for public-sector decision-making, since it directly informs whether the agent system would even be an economically viable option for a given task \citep[independent of considerations of accountability or legitimacy; cf.][]{schmitzMoralAgencyFramework2025}.

\paragraph{Fairness:}
Another metric type -- which is much more challenging to define -- is \emph{fairness} \citep{wachterWhyFairnessCannot2021}. At its core, fairness in AI systems is about ensuring that different groups (e.g., people with different genders, ages, or ethnicities) are treated fairly. Of course, this definition strongly depends on how we define `fairly', which is a question that has puzzled philosophers for centuries \citep{binnsFairnessMachineLearning2018}. The concrete definition of fairness depends on the context of not just the AI system, but the socio-technical system in which it is embedded \citep{selbstFairnessAbstractionSociotechnical2019}. Sometimes, fairness should be operationalised as equal performance between different groups \citep[e.g., in the case of resume classification;][]{hardtEqualityOpportunitySupervised2016}, whereas in other scenarios it is more important the AI system achieves some threshold of performance for all groups \citep[e.g., in medical diagnostics;][]{mittelstadtUnfairnessFairMachine2024}. See \citet{mitchellAlgorithmicFairnessChoices2021} for an overview of fairness metrics.

Regardless of the thorniness of its definition, measuring fairness is essential for public-sector systems. Fairness and accountability are fundamental values across many public sectors \citep{jorgensenPublicValuesTheir2007,vanderwalWhatsValuedMost2008}. All applications will have context-specific statutory requirements and norms, where fairness metrics play an important role in human evaluation \citep{wachterWhyFairnessCannot2021}. Having a norm of reporting context-specific fairness metrics is a necessary step for moving towards organisational accountability \citep{mittelstadtPrinciplesAloneCannot2019}. While it is beyond the scope of this paper to evaluate if benchmarks use \emph{appropriate} fairness metrics given the ethical and legal context, we will evaluate \emph{whether} benchmarks use fairness metrics. 

An example of a paper measuring fairness is \emph{MAPS} \citep{hofmanMAPSMultilingualBenchmark2025}, which reports agent performance across 11 different languages, which enables analysis of disparities and challenges.

\paragraph{Other metrics:}
Fairness and cost are not the only relevant additional metrics to report. For instance, \citet{kapoorAIAgentsThat2025} mentions \emph{robustness} (in the sense of robustness to distribution shifts) as important for agent applications. Another kind of robustness is \emph{adversarial robustness} \citep{buzhinskyMetricsMethodsRobustness2023}, which is whether small perturbation to the input can lead to different decisions. Systems with poor adversarial robustness will be prone to attacks and be gameable, which can hurt individual fairness by introducing shortcuts. 

This requires carefully considering which changes can affect the agent once deployed. Since LLMs have inherent randomness, robustness is important to measure \citep{benderDangersStochasticParrots2021,khoujaLINGOLYTOODisentanglingReasoning2025}. However, in preliminary experiments, we found high variation in definitions of robustness, which made it infeasible to include specific guidance. 

Furthermore, each specific public-sector use case will have specific socio-technical constraints that should be accounted for \citep{selbstFairnessAbstractionSociotechnical2019}. The two metrics we focus on in this paper (`cost' and `fairness' metrics) should therefore be seen as a starting point rather than a comprehensive checklist. 

\section{Methods} \label{sec:methods}
\begin{figure}
    \centering
    \includegraphics[width=1.0\linewidth]{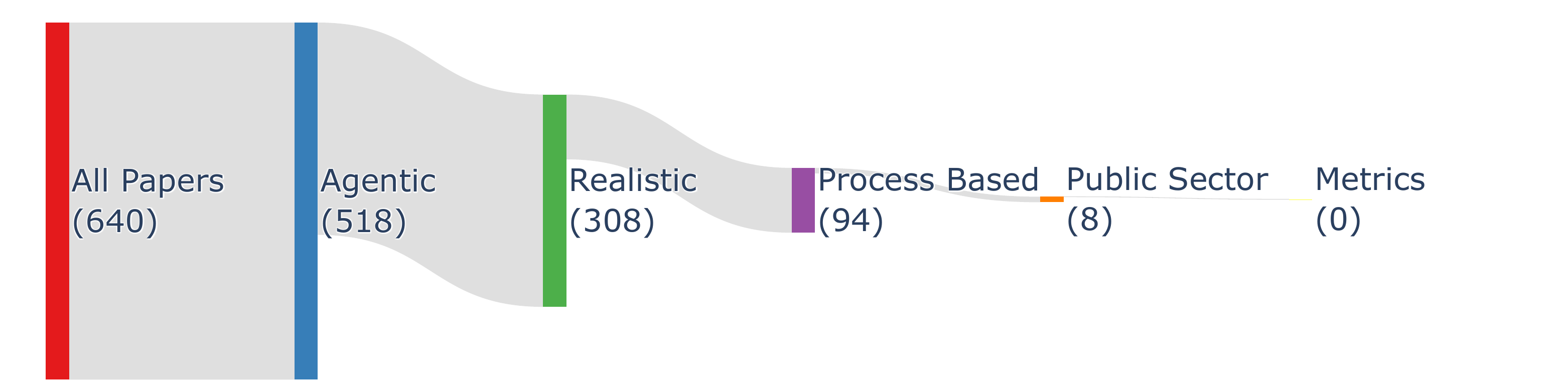}
    \caption{Flow of how well the surveyed papers meet our criteria (see \secref{sec:criteria}). No single paper meets all of the criteria, with the most challenging categories being public-sector relevance and relevant metrics.}
    \label{fig:paper-sankey}
\end{figure}

Here, we describe how we survey the literature to construct an overview of the state of public-sector relevant agent benchmarks. We implement an iterative LLM-assisted distant reading methodology \citep{hsuCHIMELLMassistedHierarchical2024}. First, we conduct a keyword search on ArXiv as well as various machine learning conferences (ICLR, NeurIPS, and ACL Anthologies) using the keywords ``LLM'', ``Benchmark'', and ``Agent''. We then deduplicate the papers using fuzzy matching on the titles using a threshold of 90\%, which removes 29 duplicates \citep{bachmannRapidfuzzRapidFuzzRelease2025}. This results in 1312 unique papers. We download the full text of all these papers from ArXiv or the relevant proceedings. On a high level, we iteratively refine the classification of these papers using a combination of LLM and expert labelling.

\subsection{LLM labelling and operationalisation}
The principal step of our iterative flow is to operationalise the criteria discussed in \secref{sec:criteria} in a way that can be processed by an LLM. Here, we use zero-shot structured data extraction \citep{chenEnhancingFunctioncallingCapabilities2025}, i.e., defining a structured schema (using JSON), which constrains the outputs of the LLM. Crucially, this allows us to define the targets (i.e., the criteria) in natural language. 

To increase performance, we use a chain-of-thought-inspired workflow \citep{weiChainofthoughtPromptingElicits2022}. This means that before the LLM outputs a label for the criteria, it writes a brief analysis which provides reasoning traces that we can use to qualitatively assess alignment \citep{brailasArtificialIntelligenceQualitative2025}. The exact structure of the data model can be seen in Appendix \ref{app:prompts}. To create a fully specified response set, we allow each criterion to be labelled as `Yes', `No', or `Unknown' \citep{guerdanValidatingLLMasajudgeSystems2025}. Note, that in the first iteration, we employ naive descriptions and system prompts to kick off the iterative refinement.

We use the open weights \texttt{gpt-oss-120b} for the screening with respect to whether the paper implements a benchmark and adheres to our definition of agenticness (see \secref{sec:agents-background}). Since the model is open-weights it increases reproducibility relative to closed models \citep{rogersClosedAIModels2023}. We use a temperature of 0 to ensure reproducibility further.

\subsection{Expert labelling}
To ensure the reliability of the LLM labelling, we manually annotate a subset of papers. We select the subset based on the first iteration of the LLM labelling such that we have at least two papers per category (e.g., two public sector/non-public sector, measures cost/doesn't measure cost, realistic/not realistic etc.) This results in a total of 54 sampled papers. We equally divide the sampled papers between the authors such that each paper has two assigned labellers. This allows us to calculate inter-rater reliability on a per-attribute basis \citep{krippendorffComputingKrippendorffsAlphareliability2011} as well as increase the reliability of the labelled data by selecting only items with agreement \citep{guerdanValidatingLLMasajudgeSystems2025}. 

We label the papers based on a survey with questions corresponding to each criteria as well as two screening attributes: whether the paper contributes a novel benchmark and whether the paper measures an agent as per our definition (see \secref{sec:agents-background}). The full survey can be found in Appendix \ref{app:survey}. We first pilot the survey to ensure we have rough agreement on the constructs within the labelling group.

The author group went through three iterations of refining definitions before running the final LLM-assisted analysis. At the end of each iteration, we manually analysed disagreements and did conflict resolution loosely following the steps outlined in \citet{oortwijnInterraterDisagreementResolution2021}.

This dataset allows us to design a prompt that accurately captures the concepts. Our initial prompt was exactly the instructions used for annotation.
Following \citet{guerdanValidatingLLMasajudgeSystems2025}, we design a prompt that meets two criteria: high alignment with expert labellers and similar selection rates. This ensures that we capture both similar judgements and similar rate statistics. We iterate the prompt by adding and modifying inclusion and exclusion criteria until we reach sufficient alignment.

We show the inter-rater reliability statistics in Table \ref{tab:interrater}. We report both Krippendorf's alpha ($\alpha_K$) and the \emph{positive rate}, i.e., the fraction of papers rated as belonging to the given category. Conventionally, an $\alpha_K>60\%$ is seen as moderate reliability \citep[cf.][]{landisMeasurementObserverAgreement1977}. However, one should be careful about dogmatic interpretations \citep{gigerenzerMindlessStatistics2004}---especially with the complex constructs in this paper \citep{wongCrossreplicationReliabilityEmpirical2021}. Still, the human-LLM reliability is generally high (with the exception of process-based). For the rates, it's important to note that the LLM has generally higher rates than the humans.

% % Old table:
% \begin{table}
% \small
% \centering
% \caption{Krippendorf's $\alpha$ and positive rate for LLMs and annotators}
% \label{tab:interrater_old}
% \begin{tabular}{llccccccc}
% \toprule
% \multirow{2}{*}{} & \multirow{2}{*}{} & \textbf{Agentic} & \textbf{Process} & \textbf{Realistic} & \textbf{Public} & \textbf{Cost} & \textbf{Fairness} \\
% & & & \textbf{Based} & & \textbf{Sector} & & \\
% \midrule
% \multirow{2}{*}{$\alpha_{K}$} & Human-Human & 38.2\% & 63.5\% & 27.8\% & 53.3\% & 52.8\% & 37.6\% \\
% & Human-LLM & 74.1\% & 74.7\% & 39.1\% & 65.6\% & 87.1\% & 47.5\% \\
% \cmidrule{1-8}
% \multirow{2}{*}{Rate} & Human & 56.5\% & 31.0\% & 42.9\% & 2.9\% & 15.6\% & 6.1\% \\
% & LLM & 55.1\% & 34.6\% & 32.1\% & 10.3\% & 14.1\% & 9.0\% \\
% \bottomrule
% \end{tabular}
% \end{table}

\begin{table}
\small
\centering
\caption{Krippendorf's $\alpha$ and positive rate for LLMs and annotators}
\label{tab:interrater}
\begin{tabular}{llccccc}
\toprule
\multirow{2}{*}{} & \multirow{2}{*}{} & \textbf{Process} & \textbf{Realistic} & \textbf{Public} & \textbf{Cost} & \textbf{Fairness} \\
& & \textbf{Based} & & \textbf{Sector} & & \\
\midrule
\multirow{2}{*}{$\alpha_{K}$} & Human-Human & 62.6\% & 73.3\% & 73.2\% & 64.4\% & 47.8\% \\
& Human-LLM & 29.2\% & 57.7\% & 79.3\% & 100.0\% & 78.5\% \\
\cmidrule{1-7}
\multirow{2}{*}{Rate} & Human & 8.8\% & 46.9\% & 23.5\% & 8.1\% & 8.8\% \\
& LLM & 27.3\% & 63.6\% & 15.2\% & 6.1\% & 9.1\% \\
\bottomrule
\end{tabular}
\end{table}

\section{State of the Literature} \label{sec:results}
\begin{figure}[htbp]
    \centering
    \begin{subfigure}[b]{0.55\textwidth}
        \centering
        \includegraphics[width=\textwidth]{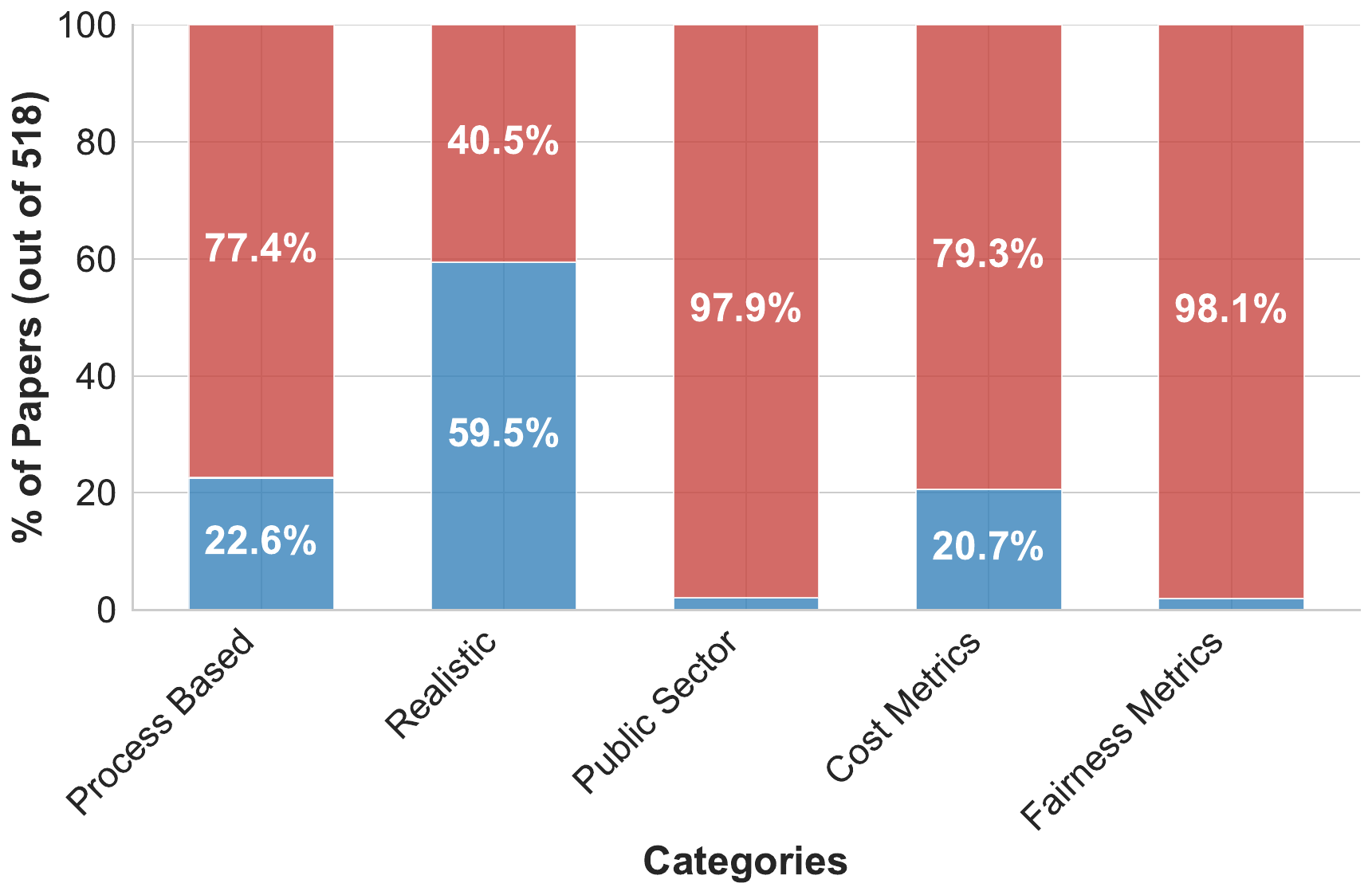}
        \caption{Unconditional distribution of the different criteria. The rarest criteria are public sector relevance and fairness.}
        \label{fig:stacked}
    \end{subfigure}
    \hfill  % Creates horizontal spacing between subfigures
    \begin{subfigure}[b]{0.40\textwidth}
        \centering
        \includegraphics[width=\textwidth]{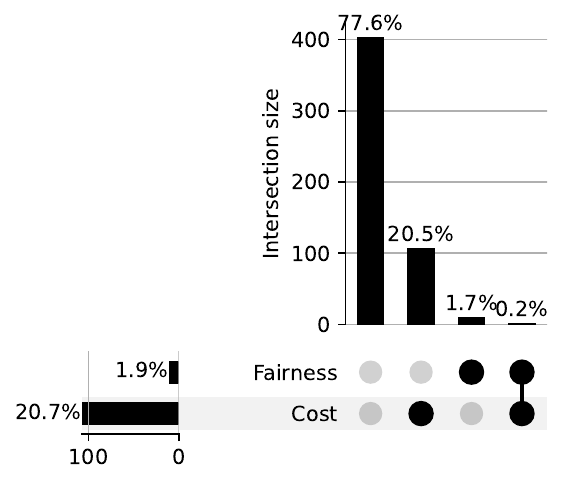}
        \caption{Distribution of reported metrics. The rarest metric is fairness, and 77.6\% of papers report none of our metrics. }
        \label{fig:upset-metrics}
    \end{subfigure}
    \label{fig:mainfig}
\end{figure}
In this section, we analyse the results of our LLM-assisted literature review. First, we analyse the high-level patterns our review has uncovered. Then, we carefully analyse the few papers that best meet our criteria. 

\paragraph{High-level pattern: No complete benchmark exists}
We show the overall flow of our criteria in Fig. \ref{fig:paper-sankey}. There are several things to note about the state of PSO-relevant agent benchmarks. The first and most important finding is that no single benchmark meets all of our criteria. 

As the figure illustrates, the two main limiting factors are reporting all relevant metrics and being relevant to the public sector. Fig. \ref{fig:upset-metrics} shows that only one paper \citep{wangXWebAgentBenchMultilingualInteractive2025} reports both cost and fairness  --- regardless of being public-sector relevant. 77.6\% of the papers report neither fairness nor cost (see \secref{sec:metrics}). Of these, fairness is the least reported (1.9\% of papers) with cost being reported by 20.7\% (see also Fig. \ref{fig:stacked}). As argued in \secref{sec:metrics}, reporting these metrics is important for public-sector decision-makers to make informed decisions about the real-world usability of the benchmarks. We will discuss the specific cases of fairness and public-sector relevant benchmarks in the following sections. 

\paragraph{Fairness is not integrated:}
As evident from Fig. \ref{fig:upset-metrics}, only 1.9\% of papers report fairness metrics. Most of these papers are exclusively focused on fairness evaluation. For instance, \citet{turSafeArenaEvaluatingSafety2025},\citet{zhengALIagentAssessingLLMs2024}, and \citet{hofmanMAPSMultilingualBenchmark2025} all explicitly evaluate safety and alignment, while \citet{liBIASINSPECTORDetectingBias2025} evaluates algorithmic bias. Only \citet{schmidgallAgentClinicMultimodalAgent2025} and \citet{shenLawReasoningBenchmark2025} integrate fairness metrics in non-fairness-focused benchmarks. 

As previously argued, measuring fairness is an essential (but fraught) requirement for deploying agentic solutions in the public sector (see \secref{sec:metrics}). Unforeseen failures are the norm, not the exception. It is thus imperative that future benchmarks incorporate fairness measures regardless of their focus.
        
\paragraph{Few public-sector-specific benchmarks:}
Finally, a core limitation is that there are few public-sector relevant benchmarks. As seen in Fig. \ref{fig:paper-sankey}, there are only eight process-based benchmarks uniquely relevant for the public sector per our definitions (see \secref{sec:criteria}). These benchmarks focus on judicial administration \citep{heAgentsCourtBuildingJudicial2024,shenLawReasoningBenchmark2025}, climate finance \citep{vaghefiAIClimateFinance2025}, logistics \citep{sunVTSLLMDomainadaptiveLLM2025,laiUSTBenchBenchmarkingDissecting2025}, and security \citep{xiangGuardAgentSafeguardLLM2025}.

While the above benchmarks cover a broad range of use cases, they are far from spanning the full breadth of administration tasks within public-sector organizations. Ideally,a plethora of public-sector relevant benchmarks should be available to inform implementation efforts. While the lack of existing benchmarks presents a challenge, it also provides an opportunity for researchers to develop these benchmarks. 

This is not to say that benchmarks must be specific to the public sector to have informative value. Benchmarks like WorkArena \citep{drouinWorkArenaHowCapable2024} and FOFO \citep[format following;][]{xiaFOFOBenchmarkEvaluate2024} , which tests the ability of language models to produce text of specific formats, measure important abstract capabilities that are relevant in the public sector. Still, the stringent requirements and process-based flow in public-sector organisations might be particularly challenging for LLM-based agents; we will only know when we start measuring them.

\section{Conclusion} \label{sec:conclusion}
The safe adoption of LLM-based agents in public-sector institutions is a sociotechnical challenge spanning technology and organisation \citep{loweFullstackAlignmentCoaligning2025}. A precondition for its success is robust, contextually valid technical evaluation of these systems against the public sector's unique requirements \citep{schmitzMoralAgencyFramework2025}. We derive here that candidate LLM agent benchmarks must therefore, at a minimum, live up to a set of criteria to be relevant for the public sector. We propose these criteria as a \emph{signal-strength rubric}: the more criteria a benchmark meets, the more informative it is for public-sector adoption decisions, and gaps indicate where benchmark scores provide weak evidence about real-world deployment. Concretely, benchmarks should be \emph{process-based} (measuring \emph{interdependent subtasks} that reflect how work unfolds in practice), \emph{realistic} (measuring \emph{real tasks with realistic data}), focused on \emph{public sector-specific} tasks (i.e., tasks actually performed in public-sector organisations), and report public-sector relevant \emph{metrics}, including \emph{cost} and \emph{fairness} alongside other context-dependent measures.

Using an LLM-assisted literature analysis technique, we survey more than 1,300 papers against this rubric. The results are sobering. No single benchmark meets all of the criteria. The largest gaps concern public-sector specificity and comprehensive metric reporting: only one benchmark reports both cost and fairness (with fairness rarely integrated), and there are few public-sector-specific benchmarks. These gaps risk encouraging overconfidence in agents on the basis of benchmark results that are not well-matched to public-sector constraints and trade-offs.

While our findings highlight vast public-sector gaps in the current benchmark literature, our findings also provide a call to action. We call on the academic community to: (1) develop public-sector-specific benchmarks, (2) integrate more relevant metrics, including fairness, and (3) prioritise ecological validity in benchmark design. In parallel, public-sector organisations should assess any benchmarks they rely on against these criteria, and treat missing dimensions as limits on applicability rather than as reassurance. If this call is heeded, public-sector organisations around the world will be in a better position to responsibly integrate LLM agents for the benefit of all citizens.

% The safe adoption of LLM-based agents in public-sector institutions is a sociotechnical challenge spanning technology and organisation \citep{loweFullstackAlignmentCoaligning2025}. A precondition for its success is robust, contextually valid technical evaluation of these systems against the public sector's unique requirements \citep{schmitzMoralAgencyFramework2025}. We derive here that candidate LLM agent benchmarks must therefore, at a minimum, live up to a set of criteria to be relevant for the public sector. We posit that they should be \emph{realistic}, \emph{process-based}, focus on \emph{public sector-specific} tasks, and report public-sector relevant \emph{metrics}, including \emph{cost} and \emph{fairness}. Using an LLM-assisted literature analysis technique, we survey more than 1,300 papers for these criteria. The results are sobering. No single benchmark meets all of the criteria. Specifically, only one benchmark reports relevant metrics (with fairness not being integrated), and there are few public-sector-specific benchmarks. 

% While our findings highlight vast public-sector gaps in the current benchmark literature, our findings also provide a call to action. We call on the academic community to: (1) develop public-sector-specific benchmarks, (2) integrate more relevant metrics, including fairness, and (3) prioritise ecological validity in benchmark design. If this call is heeded, public-sector organisations around the world will be in a better position to responsibly integrate LLM agents for the benefit of all citizens.

\section*{Limitations}
Despite our extensive stratified annotation strategy, it is challenging to ensure coverage across all categories. Since few papers live up to, e.g., public-sector relevance, we might not cover all types of papers in our manual annotations. However, since we carefully match the rates and agreement, we are confident in our overall conclusions.

As noted in \secref{sec:methods}, we have imperfect agreement between the LLM classifications and human annotations. However, since the annotators are generally more conservative than the LLM (see Table \ref{tab:interrater}) and our core claims are about the non-existence of complete benchmarks, we are nevertheless confident in the directionality of our conclusions.

\begin{ack}
We thank the members of the Digital State Forum for their helpful discussion. JR was supported by the Engineering and Physical Sciences Research Council [Grant Number EP/W524311/1]. JB was supported by the Federal Ministry of Research, Technology, and Space of Germany [Grant Number 16DII131]. CS and JB acknowledge funding by the Hertie School’s program ``AI and Data Science for the Public Sector'', funded by the Dieter Schwartz Foundation. KK was supported by the Clarendon Fund.
\end{ack}

\bibliographystyle{apalike}
\bibliography{references}

%%%%%%%%%%%%%%%%%%%%%%%%%%%%%%%%%%%%%%%%%%%%%%%%%%%%%%%%%%%%

\appendix

\section{Literature Review}
\subsection{Selection Methodology}
We describe our selection methodology through three main conditions that guided the inclusion of theories in the benchmark framework.

A theory was considered \emph{well-established and notable} if it was introduced in main publication(s) with over 300 citations.

A theory was regarded as \emph{salient in research on public-sector AI use} if it either (a) explicitly focuses on digital systems in public administration or bureaucracy, or (b) has been applied to the AI/digitalisation of the public sector in multiple peer-reviewed studies, government reports, or notable policy analyses.

Finally, a theory was considered \emph{robust and distinct} if it offers a coherent conceptual structure that applies across contexts and is not reducible to other models in the list. It should capture a mechanism or dimension (e.g., discretion, task structure, adoption process) that is analytically distinct and has been operationalised in at least two different empirical or policy applications.

\subsection{Selected Works}
For an overview of selected works, see Table \ref{tab:detailed_theory_table}.

% Requires \usepackage{booktabs} and \usepackage{amssymb} for \checkmark
% Requires \usepackage{booktabs}, \usepackage{tabularx}, \usepackage{array}
\renewcommand{\arraystretch}{1.4}  % comfortable row spacing

% Preamble (add these packages)
% \usepackage{booktabs}
% \usepackage{tabularx}
% \usepackage{threeparttable}
% \usepackage{array}
% \usepackage{ragged2e}

\begin{table}[ht]
\centering
\begin{threeparttable}
\caption{Canonical sources, citation counts, public-sector AI applications, and distinctiveness of the selected theories/models}
\label{tab:detailed_theory_table}
\begin{tabularx}{\textwidth}{
  >{\raggedright\arraybackslash}X
  >{\raggedright\arraybackslash}X
  >{\raggedright\arraybackslash}X
  >{\raggedright\arraybackslash}X
  >{\raggedright\arraybackslash}X}
\toprule
\textbf{Model / Theory} &
\textbf{Canonical source} &
\textbf{C1: \newline Citations}\tnote{a} &
\textbf{C2: Notable applications in public-sector AI} &
\textbf{C3: Distinct conceptual contribution} \\
\midrule

Theory of automation &
\citet{acemogluSkillsTasksTechnologies2011} &
6957 &
\citet{hashemMappingPotentialGenerative2025}; \newline \citet{straubAIBureaucraticProductivity2024}; \newline \citet{acemogluSimpleMacroeconomicsAI2024} &
Task-substitution vs.\ augmentation framing \\[.4em]

Street-, Screen-, System-level bureaucracy &
\citet{lipskyStreetlevelBureaucracy30th2010};\newline \citet{bovensStreetlevelSystemlevelBureaucracies2002} &
29151;\newline 1403 &
Inherently Public Sector Digitalisation; \newline e.g.\ \citet{soaresScreenlevelBureaucratsAge2024} &
Conceptualisation of discretional shift from humans to systems \\[.4em]

Digital-Era Governance &
\citet{dunleavyDigitalEraGovernance2006} &
1459 &
Inherently Public Sector Digitalisation &
Holistic framework for service transformation with focus on process-level digitalisation \\[.4em]

Artificial Discretion &
\citet{youngArtificialDiscretionTool2019}\tnote{b} &
303 &
Inherently Public Sector Digitalisation &
Extension of bureaucratic discretion literature to automated decision-making \\[.4em]

Government by Algorithm &
\citet{engstromGovernmentAlgorithmArtificial2020} &
308 &
Inherently Public Sector Digitalisation &
Extension of governance by expert system proposed by \citet{simonAdministrativeBehaviorStudy1947} \\[.4em]

Diffusion of Innovations &
\citet{madanAIAdoptionDiffusion2023} &
162601 &
\citet{madanAIAdoptionDiffusion2023}; \newline 
\citet{devriesDiffusionAdoptionPublic2018}; \newline \citet{bokhariArtificialIntelligencebasedTechnologicaloriented2022} &
Technology adoption determinants (relative advantage, compatibility, trialability) \\[.4em]

TOE Framework &
\citet{tornatzkyProcessesTechnologicalInnovation1990} &
2727\tnote{c} &
\citet{neumannExploringArtificialIntelligence2024}; \newline \citet{paivaExploringDriversAI2024} &
Widely used tri-dimensional lens (technology, organisation, environment) \\
\bottomrule
\end{tabularx}

\begin{tablenotes}
\item[a] As of July 2025.
\item[b] One of several competing ``canonical'' references.
\item[c] Count of \citet{oliveiraLiteratureReviewInformation2011}, as original count is unavailable.
\end{tablenotes}
\end{threeparttable}
\end{table}

\section{Prompts} \label{app:prompts}
\subsection{Text Prompts}
\begin{lstlisting}[language=Python]
BENCHMARK_SCREENING_SYSTEM_PROMPT = """
You are a SOTA literature classification agent. You are given a paper title and abstract. Your recall and precision has been evaluated as surpassing 99%.
You only label papers as having agentic benchmarks if they introduce a novel benchmark dataset for evaluating agents. Invalid if the paper a) introduces a method and not a benchmark, b) only evaluates on existing benchmarks, c) doesn't have a benchmark (and not a method) as the primary contribution.
"""

AGENT_SCREENING_SYSTEM_PROMPT = """
You are a SOTA literature classification agent with 99%+ accuracy.

Label papers as agentic AI if they involve:
- AI systems that autonomously use tools, interact with environments, or perform multi-step tasks
- Benchmarking core autonomous capabilities (planning, tool use, environmental interaction, format-following for agent applications)
- Comparing agent vs non-agent approaches to autonomous tasks
- Multi-agent systems where individual agents use tools or interact with environments

NOT agentic AI:
- Multi-agent systems where agents only communicate/debate without individual tool use
- Bias evaluation, dataset creation, or general LLM assessment without autonomous behavior focus
- Domain-specific Q&A or classification tasks
- Inference methods or prompting techniques alone

Key test: Does it evaluate or build systems that act autonomously beyond simple text generation?
""".strip()

METRIC_SYSTEM_PROMPT = """You are a specialized literature classification agent for AI benchmark evaluation papers. You classify whether benchmarks measure Cost and Fairness.%, and Robustness.
% Robustness evaluation focuses on measuring how agent performance varies across different conditions. Specifically, look for benchmarks that:
% 1. **Robustness between runs**: Testing consistency across multiple runs on the same task.
% Look for benchmarks that **measure** how performance changes across these conditions, not just benchmarks that **include** varied conditions. The measurements will be different from performance metrics.
"""

VALIDITY_SYSTEM_PROMPT = """You are a SOTA literature classification agent. You are given a paper title and abstract. Your recall and precision has been evaluated as surpassing 99%.

CRITICAL DISTINCTIONS:

For "realistic_tasks":
DIRECT WORK OUTPUTS: Tasks where humans are employed to produce the actual deliverable (diagnosis, code, analysis report)
AI TESTED ON REAL WORK TASKS: AI generating actual work outputs humans are paid to produce (e.g., code for real applications)
AI PERFORMANCE EVALUATION: Testing AI capabilities in simulated/controlled environments, even if professionally relevant
TOOL/SYSTEM EVALUATION: Testing AI frameworks, architectures, or tool coordination
ACADEMIC/GAME SCENARIOS: Abstract reasoning, board games, simulations designed for AI evaluation

CRITICAL TEST: "Does this produce actual work deliverables that someone would pay for?" not "Is this in a professional domain?"

For "explicit_subtasks" and "interdependent_subtask_process":
FOCUS ON INTERMEDIATE OUTPUTS WITH REAL VALUE:

RECOGNIZE AS SUBTASKS WHEN:
- Intermediate outputs have standalone utility that could be used independently
- Work phases represent natural breakpoints for review or handoff
- Phases require meaningfully different skills or approaches
- Domain professionals would naturally recognize distinct work phases

DO NOT COUNT AS SUBTASKS:
- AI system processing stages or internal coordination
- Simple reasoning steps within single deliverables
- Sequential operations that are just implementation details
- Complex reasoning that produces only one final deliverable

KEY TEST: "Does the intermediate output have independent value that could be handed off or reviewed?"

MAINTAIN BALANCE:
- Don't dismiss genuine work phases with valuable intermediate outputs
- Don't count every processing step or AI complexity as subtasks
"""

\end{lstlisting}

\subsection{Classification Schemas}
We do structured data extraction using JSON schemas \citep{pezoaFoundationsJSONSchema2016} as popularised through function calling \citep[e.g.,][]{kimLLMCompilerParallel2024}. See the definitions below.

% Generated JSON schemas formatted for minted
% Include this file in your LaTeX document with \input{minted_schemas.txt}

\subsubsection*{IsAboutAgenticAI}
\begin{lstlisting}[language=json]
{
  "$defs": {
    "Criteria": {
      "properties": {
        "analysis": {
          "description": "1 sentence extremely concise analysis of whether the paper lives up to the criteria",
          "title": "Analysis",
          "type": "string"
        },
        "valid": {
          "enum": [
            "Yes",
            "No",
            "Unknown"
          ],
          "title": "Valid",
          "type": "string"
        }
      },
      "required": [
        "analysis",
        "valid"
      ],
      "title": "Criteria",
      "type": "object"
    }
  },
  "description": "Analyzes whether a paper concerns agentic AI: individual autonomous systems, evaluation of agent behaviors/capabilities, agent frameworks, or comparisons involving agent approaches.",
  "properties": {
    "is_agentic_ai": {
      "$ref": "#/$defs/Criteria",
      "description": "Does the paper evaluate, build, or compare agent behaviors or autonomous capabilities?"
    }
  },
  "required": [
    "is_agentic_ai"
  ],
  "title": "IsAboutAgenticAI",
  "type": "object"
}
\end{lstlisting}

\subsubsection*{IntroducesAgentBenchmark}
\begin{lstlisting}[language=json]
{
  "$defs": {
    "Criteria": {
      "properties": {
        "analysis": {
          "description": "1 sentence extremely concise analysis of whether the paper lives up to the criteria",
          "title": "Analysis",
          "type": "string"
        },
        "valid": {
          "enum": [
            "Yes",
            "No",
            "Unknown"
          ],
          "title": "Valid",
          "type": "string"
        }
      },
      "required": [
        "analysis",
        "valid"
      ],
      "title": "Criteria",
      "type": "object"
    }
  },
  "description": "Analyzes whether a paper actually introduces a novel benchmark for evaluating agents. The benchmark must be named in the abstract or title.",
  "properties": {
    "introduces_agent_benchmark": {
      "$ref": "#/$defs/Criteria",
      "description": "Does the paper introduce a novel benchmark dataset for evaluating agents? Invalid if the paper a) is not about agents, b) does not introduce a benchmark (but rather a framework etc.) or c) only uses existing benchmarks. Name the benchmark in the analysis."
    }
  },
  "required": [
    "introduces_agent_benchmark"
  ],
  "title": "IntroducesAgentBenchmark",
  "type": "object"
}
\end{lstlisting}

\subsubsection*{ValidAgentBenchmark}
\begin{lstlisting}[language=json]
{
  "$defs": {
    "Criteria": {
      "properties": {
        "analysis": {
          "description": "1 sentence extremely concise analysis of whether the paper lives up to the criteria",
          "title": "Analysis",
          "type": "string"
        },
        "valid": {
          "enum": [
            "Yes",
            "No",
            "Unknown"
          ],
          "title": "Valid",
          "type": "string"
        }
      },
      "required": [
        "analysis",
        "valid"
      ],
      "title": "Criteria",
      "type": "object"
    }
  },
  "description": "Evaluates whether a paper describes a benchmark that is specifically relevant for public sector applications",
  "properties": {
    "is_public_sector_specific": {
      "$ref": "#/$defs/Criteria",
      "description": "Does the benchmark focus on tasks unique to government administration rather than general capabilities applicable across many sectors? (do not include healthcare)"
    },
    "realistic_tasks": {
      "$ref": "#/$defs/Criteria",
      "description": "Do the benchmark tasks produce actual work deliverables that humans are employed to create? Focus on whether these generate real work outputs someone would pay for, not just professional domain relevance. Include AI generating actual work products (e.g., functional code for applications). Exclude AI performance evaluations in simulated environments, tool/system assessments, or academic scenarios designed primarily for AI capability testing."
    },
    "real_world_data": {
      "$ref": "#/$defs/Criteria",
      "description": "Does the benchmark use real-world data?"
    },
    "explicit_subtasks": {
      "$ref": "#/$defs/Criteria",
      "description": "Would these tasks naturally involve multiple distinct work phases that produce intermediate outputs with independent value? Focus on whether intermediate outputs have standalone utility and could be handed off to others. Include domain-standard work breakdowns where phases require meaningfully different skills. Exclude AI system processing stages, simple reasoning steps, or coordination complexity without meaningful intermediate outputs."
    },
    "interdependent_subtask_process": {
      "$ref": "#/$defs/Criteria",
      "description": "Do these tasks require producing meaningful intermediate work products that serve as essential inputs for subsequent phases? Focus on intermediate outputs that have independent value and represent natural breakpoints where work could be reviewed before proceeding. Exclude AI internal processing, simple sequential steps, or cases where intermediate outputs are just reasoning artifacts."
    }
  },
  "required": [
    "is_public_sector_specific",
    "realistic_tasks",
    "real_world_data",
    "explicit_subtasks",
    "interdependent_subtask_process"
  ],
  "title": "ValidAgentBenchmark",
  "type": "object"
}
\end{lstlisting}

\subsubsection*{BenchmarkEvaluationMetrics}
\begin{lstlisting}[language=json]
{
  "$defs": {
    "Criteria": {
      "properties": {
        "analysis": {
          "description": "1 sentence extremely concise analysis of whether the paper lives up to the criteria",
          "title": "Analysis",
          "type": "string"
        },
        "valid": {
          "enum": [
            "Yes",
            "No",
            "Unknown"
          ],
          "title": "Valid",
          "type": "string"
        }
      },
      "required": [
        "analysis",
        "valid"
      ],
      "title": "Criteria",
      "type": "object"
    }
  },
  "description": "Evaluates whether a benchmark includes specific evaluation metrics that are relevant for agentic AI.",
  "properties": {
    "evaluates_cost": {
      "description": "Does the benchmark measure computational resources, time, or financial costs of task execution?",
      "enum": [
        "Yes",
        "No",
        "Unknown"
      ],
      "title": "Evaluates Cost",
      "type": "string"
    },
    "evaluates_fairness": {
      "description": "Does the benchmark assess bias, disparate impacts, or equitable outcomes across different demographic groups?",
      "enum": [
        "Yes",
        "No",
        "Unknown"
      ],
      "title": "Evaluates Fairness",
      "type": "string"
    },
    "evaluates_robustness": {
      "$ref": "#/$defs/Criteria",
      "description": "Does the benchmark report separate measurements of how agent performance varies across different conditions, such as multiple runs, distribution shifts, or varying task scenarios? Explicitly mention the exact metric used."
    }
  },
  "required": [
    "evaluates_cost",
    "evaluates_fairness",
    "evaluates_robustness"
  ],
  "title": "BenchmarkEvaluationMetrics",
  "type": "object"
}
\end{lstlisting}

\section{Survey Questions} \label{app:survey}

\begin{enumerate}
    \item \textbf{Benchmark:} Does the paper actually implement and introduce a novel benchmark?
    
    \item \textbf{Agentic:} Does the benchmark concern agentic LLMs?
    
    \item \textbf{Subtasks:} Does the benchmark report the performance of sub-tasks?
    
    \item \textbf{Sequential:} Are the subtasks sequentially dependent? (Process-based)
    
    \begin{quote}
        \emph{Note: process-based = subtasks + sequential}
    \end{quote}
    
    \item \textbf{Realistic task:} Are the tasks in the benchmark realistic for the application scenario?
    
    \item \textbf{Realistic data:} Does the task use real-world data?
    
    \begin{quote}
        \emph{Note: realistic task + realistic data}
    \end{quote}
    
    \item \textbf{Public Sector:} Is the benchmark public sector relevant?
    
    \item \textbf{Metrics:} Which of the following metrics are reported? (if any)
    
    \emph{Options: Cost, Fairness}%, Robustness}
\end{enumerate}

%%%%%%%%%%%%%%%%%%%%%%%%%%%%%%%%%%%%%%%%%%%%%%%%%%%%%%%%%%%%

%%% END INSTRUCTIONS %%%

\end{document}